\documentclass[letterpaper]{IEEEtran}

\usepackage[T1]{fontenc}
\usepackage{graphicx}
\usepackage{booktabs}
\usepackage{cite}
\usepackage{url}
\usepackage{tikz}
\usepackage{hyperref}

\newcommand{\figureframe}[1]{%
  \begingroup
  \setlength{\fboxsep}{2pt}%
  \setlength{\fboxrule}{0.5pt}%
  \fbox{#1}%
  \endgroup
}

\begin{document}

\title{Pinching Antennas for ISAC: Waveguide Architectures, Solutions, and Future Directions}

\author{Qian~Gao,~\IEEEmembership{Graduate Student Member,~IEEE,}
and Ruikang~Zhong,~\IEEEmembership{Member,~IEEE,}
\thanks{Qian Gao is with the School of Electronic Engineering and Computer Science, Queen Mary University of London, London E1 4NS, U.K. (e-mail: q.gao@qmul.ac.uk).}%
\thanks{Ruikang Zhong is with the School of Information and Communication Engineering, Xi'an Jiaotong University, Xi'an 710049, China (e-mail: ruikang.zhong@xjtu.edu.cn).}%
}

\maketitle

\begin{abstract}
Integrated sensing and communication (ISAC) is a core design principle for sixth-generation (6G) wireless systems, yet fixed antenna geometry limits its performance. Pinching antenna systems (PASS) guide electromagnetic waves along dielectric waveguides and radiate them through position-adaptive antennas, allowing communication links and sensing observations to benefit from spatial reconfiguration. This article reviews PASS-enabled ISAC through single-waveguide, multi-waveguide, and segmented-waveguide architectures, comparing their feed structures, spatial degrees of freedom, operating protocols, and uplink/downlink support. We organize solution pathways around physics-consistent modeling, joint spatial-signal optimization, dynamic channel acquisition, and stable real-time control. A segmented waveguide-enabled pinching-antenna array (SWAN) case study demonstrates joint segment selection, antenna positioning, and transmit beamforming using segment-hysteresis reinforcement learning. The results illustrate how waveguide loss and per-segment aperture gain create communication-sensing tradeoffs. We conclude with research directions in joint transmit-receive operation, measurement-calibrated digital twins, and trustworthy control.
\end{abstract}

\section{Introduction}

Integrated sensing and communication (ISAC) is a defining capability for 6G, enabling the same infrastructure to deliver data and sense its surroundings~\cite{ISAC6G}. Roadside units can communicate with vehicles while detecting hazards, and factory access points can support wireless control while tracking robots and workers. These services require both efficient spectrum use and spatial awareness.

Communication favors high signal-to-interference-plus-noise ratio, user fairness, and low overhead, whereas sensing may require angular diversity, stable illumination, and accurate delay or position estimation. Conventional ISAC negotiates these requirements through waveform and digital beamforming design with a fixed aperture. This limits service in elongated regions: a fixed array can steer a beam but cannot relocate its radiation origin when users or targets move. PASS transports electromagnetic energy through a dielectric waveguide and releases it at configurable positions, adapting both the phase relationship and the location of the radiating aperture~\cite{PA}.

This characteristic makes PASS particularly appealing for ISAC. Radiating points can be placed near communication hotspots, activated from different look angles for target sensing, or distributed over several waveguides without installing a conventional phased array at every location. Joint transmit and pinching beamforming studies show how PASS can shape communication links~\cite{PASSjoint}, while waveguide division multiple access treats the guide itself as a multiplexing resource~\cite{WDMA}. User-centric positioning highlights the localization value of movable radiation points~\cite{PAISACCRLB}, and look-angle diversity shows why geometry matters for target-dependent sensing reliability~\cite{TargetDiversity}. Integrated multicast-sensing designs further indicate that PASS can support joint service objectives rather than only stronger point-to-point beams~\cite{PAMulticast}.

From a systems viewpoint, this article asks what changes when ISAC is implemented over waveguide-based, geometry-adaptive infrastructure. It explains PASS operating logic, compares single-, multi-, and segmented-waveguide architectures, maps core solution pathways, uses a SWAN case study, and concludes with future directions on digital twins, transmit-receive waveguides, and trustworthy control.

\begin{figure*}[t]
\centering
\figureframe{\includegraphics[width=0.8\textwidth]{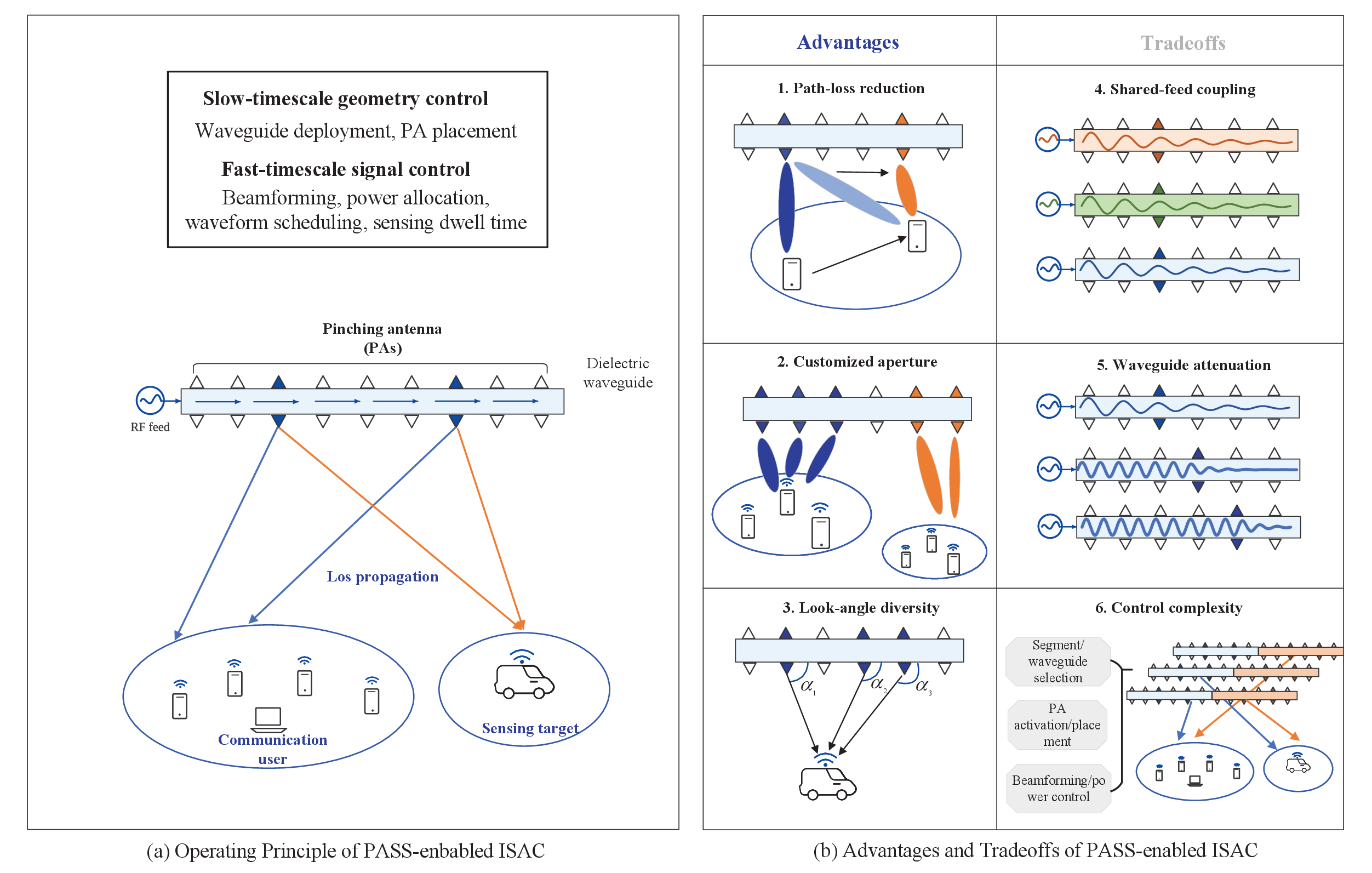}}
\caption{PASS-enabled ISAC combines guided-wave and free-space propagation. PA positions determine path loss, phase, and sensing look angle.}
\label{fig:overview}
\end{figure*}

\section{Why Pinching Antennas for ISAC?}
\label{section:2}

\subsection{Operating Principles of Pinching-Antenna-Enabled ISAC}

PASS starts from a different physical picture than a conventional antenna array. Instead of radiating all energy from a fixed panel or tower, radio-frequency energy is first launched into a dielectric waveguide. Small radiating structures, often described as pinching points or pinching antennas, are then activated at selected positions to couple part of the guided wave into free space~\cite{PA}. In this way, the network controls where radiation emerges from the infrastructure. The important consequence is that moving or selecting pinching points changes both the large-scale propagation distance and the phase profile seen by users and targets.

For ISAC, this matters since communication and sensing are both spatial services. A communication user cares about received power, interference suppression, and coverage continuity. A sensing target cares about how strongly and from which directions it is illuminated, and how the returned echo can be separated from clutter and other reflections. With PASS, the same infrastructure can place active radiating points closer to a user cluster, stretch them along a corridor to improve spatial awareness, or distribute them around an area that requires wider sensing coverage. Instead of asking a fixed array to serve all geometric situations from a single origin, the system adjusts the origin of radiation itself, as illustrated in Fig.~\ref{fig:overview}.

This new degree of freedom introduces a natural two-timescale interpretation of PASS-enabled ISAC. The slow timescale concerns geometric decisions, such as how many waveguides should be deployed and where they are located. The fast timescale concerns signal decisions, including beamforming, power allocation, waveform scheduling, and sensing dwell time. Since communication and sensing react differently to geometry, PASS can create service states that are difficult to realize with rigid apertures. This geometry-aware service shaping is the fundamental reason PASS deserves attention in the ISAC landscape.

\subsection{Advantages and Design Tradeoffs of Pinching Antennas for ISAC}

The first major advantage of PASS for ISAC is path-loss management through geometric proximity. Instead of fighting attenuation only with more power or a larger fixed array, PASS can activate radiating points nearer to a user or sensing region. This is attractive in tunnels, rail platforms, warehouse aisles, and production lines, where waveguides can follow the dominant service corridor. A second advantage is distributed aperture formation without duplicating a complete phased-array front end at every location. A third is sensing-view diversity: activating different positions over time changes the target look angle and can reduce positioning error~\cite{PAISACCRLB} or sensing outage~\cite{TargetDiversity}.

A basic single-waveguide PASS uses multiple pinching antennas sharing a common feed~\cite{PA}, limiting independent waveform control. This coupling creates tension between communication beams and sensing illumination. Scheduling can accommodate simpler service requirements; simultaneous independent control may require multiple waveguides, segmentation, or hybrid signal structures.

Waveguide loss and calibration overhead form a second tradeoff. As the infrastructure grows longer and denser, attenuation, coupling losses, and hardware mismatch across segments can accumulate. In ISAC, these effects influence not only communication rate but also echo quality and geometric consistency. Finally, the optimization burden rises quickly with scale since every additional waveguide, segment, or candidate pinching location enlarges the configuration space. The central question is therefore how to expose PASS flexibility to ISAC algorithms without overwhelming estimation, signaling, and real-time control resources.

\section{Waveguide Architectures and ISAC Operating Modes}
\label{section:3}

\begin{figure*}[t]
\centering
\figureframe{\includegraphics[width=0.8\textwidth]{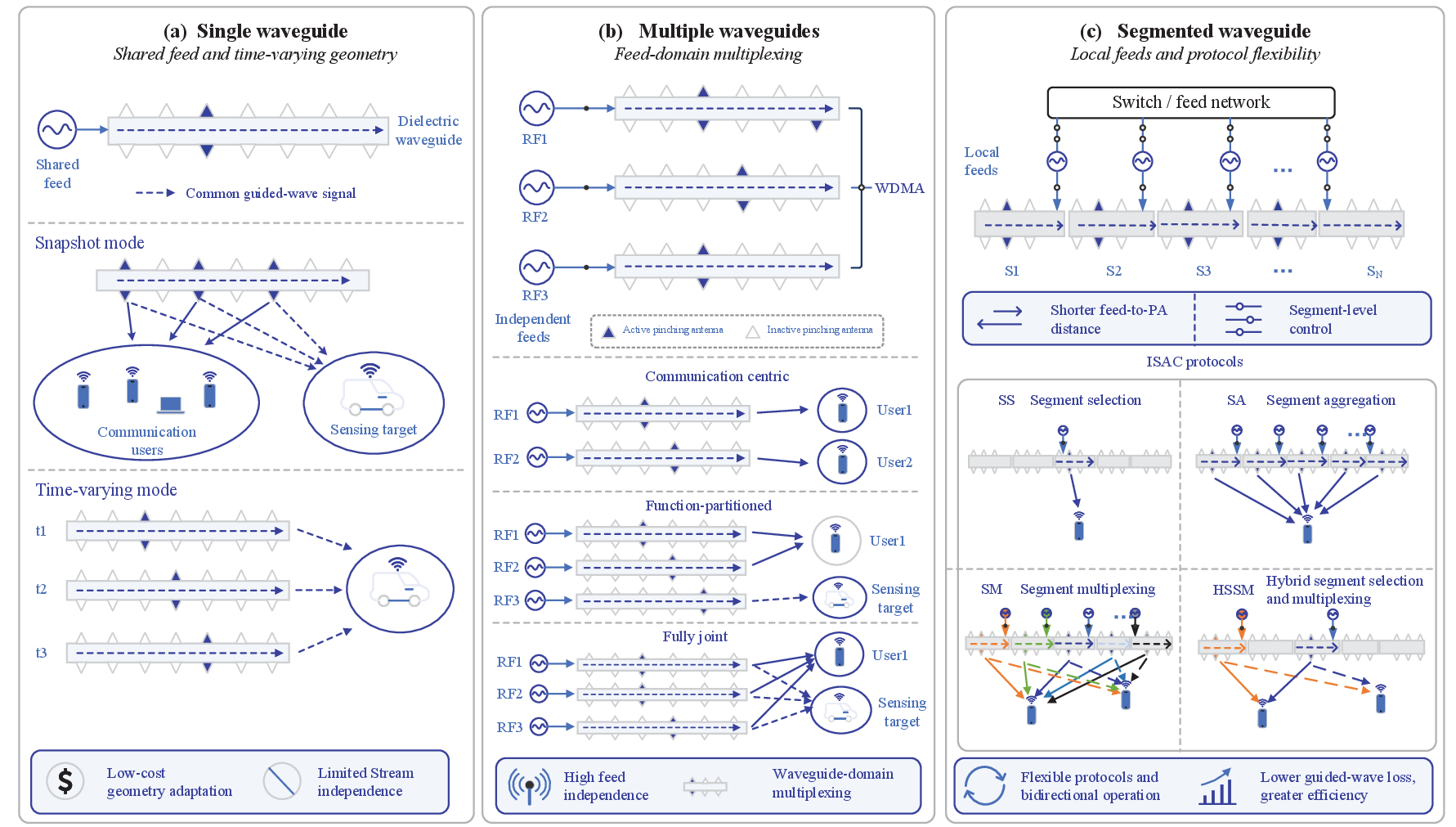}}
\caption{Waveguide architectures and operating modes for PASS-enabled ISAC. The comparison highlights the progression from shared-feed single waveguides to independent multi-waveguide resources and segmented SWAN operation.}
\label{fig:architectures}
\end{figure*}

\subsection{Single Waveguide: Shared Feed and Time-Varying Geometry}

A single waveguide is the minimum PASS implementation. One RF feed launches a waveform into the guide, and one or more pinching antennas couple that waveform into free space. The controllable variable is the radiation position; the signals emitted by antennas on the same guide remain linked through the common feed and the guided-wave phase. This architecture is therefore highly economical, but it does not provide one independent baseband stream per pinching antenna.

For ISAC, a single waveguide supports two useful modes. In a \emph{snapshot mode}, several antennas form a distributed aperture for communication and target illumination. Their positions jointly determine the free-space distance, guided-wave phase, and coherent combining pattern. In a \emph{time-varying mode}, different positions are activated across slots. The latter is particularly valuable for sensing since the same target can be observed from different look angles. Existing analysis shows that movable pinching positions can reduce the Cram\'er--Rao lower bound for user-centric positioning~\cite{PAISACCRLB}, while temporal activation diversity improves sensing reliability when radar cross section depends on look angle~\cite{TargetDiversity}.

Stream coupling favors one dominant user, multicast, or time-sharing, while limiting simultaneous independent beams for multiple users and targets. The key decisions are where and when to radiate.

\subsection{Multiple Waveguides: Feed-Domain Multiplexing}

Multiple waveguides add independent feed-domain resources. Waveguide division multiple access (WDMA) assigns a dedicated guide to each user and treats the guide itself as a radio resource~\cite{WDMA}. Pinching beamforming within each guide enhances the intended link and controls inter-user leakage, while power allocation or digital precoding coordinates the independently fed guides. This architecture turns PASS from a position-reconfigurable aperture into a hybrid beamforming system: baseband or RF-domain weights operate across guides, whereas antenna positions shape propagation within each guide~\cite{PASSjoint}.

For ISAC, independent guides can serve individual communication users while their combined radiation illuminates targets, separate sensing from data delivery, or jointly support both functions. Fully joint operation offers the most spatial flexibility but requires synchronization, calibration, and sufficient RF chains.

The architecture also exposes a practical distinction between continuous and discrete placement. Continuous optimization provides an upper bound when a pinching antenna can be positioned anywhere along a guide. A realizable design selects from finite candidate locations and must account for switching constraints. Matching, penalty, and alternating-optimization methods have been used to close this implementation gap~\cite{WDMA}. For a system view, the important point is that multi-waveguide PASS exchanges hardware simplicity for simultaneous spatial multiplexing.

\begin{table*}[htbp]
	\centering
	\caption{Solution pathways for PASS-enabled ISAC.}
	\label{tab:solutions}
	\small
	\setlength{\tabcolsep}{3pt}
	\begin{tabular}{p{0.16\textwidth}p{0.22\textwidth}p{0.25\textwidth}p{0.27\textwidth}}
		\toprule
		\textbf{Layer} & \textbf{Core question} & \textbf{Typical design variables} & \textbf{ISAC role} \\
		\midrule
		Physics-consistent modeling & How should guided-wave and free-space propagation be represented? & Feed location, PA position, waveguide loss, radiation coupling, and segment boundary & Prevents geometry choices that look favorable in free space but are inefficient after waveguide loss. \\ \hline
		Metric design and optimization & Which service objective should the geometry create? & PA positions, transmit power, beamforming weights, active guides, and active segments & Connects communication rate, sensing illumination, CRLB, outage, and hardware constraints in one joint problem. \\ \hline
		State acquisition & What state should be estimated as the aperture changes? & Position metadata, waveguide transfer response, sparse channel structure, user and target geometry & Replaces exhaustive CSI acquisition with compact, uncertainty-aware communication-sensing state inference. \\ \hline
		Online control stability & How often should topology-changing actions be updated? & Candidate pruning, hysteresis, switching penalty, slow topology control, and fast beamforming control & Keeps geometric flexibility usable without excessive switching, calibration overhead, or unstable service quality. \\
		\bottomrule
	\end{tabular}
\end{table*}

\subsection{Segmented Waveguides: Local Feeds and Protocol Flexibility}

Segmented waveguide-enabled pinching-antenna systems (SWANs) divide a long waveguide into multiple short, separately accessible segments~\cite{SWAN}. Segmentation changes the physics as well as the control architecture. Shorter PA-to-feed distances reduce in-waveguide attenuation, while local segments reduce the average PA-to-user distance. In the uplink, isolation between segments also avoids the inter-antenna re-radiation effect that can occur when several receiving antennas share one continuous guide. SWAN therefore provides a more tractable route to bidirectional multi-PA operation.

Three basic protocols define how segments contribute. Segment selection (SS) activates one favorable segment. Segment aggregation (SA) combines signals from several segments for diversity or array gain. Segment multiplexing (SM) uses separately fed segments to carry parallel streams. Their capability and hardware costs increase in the order SS, SA, and SM; reported analysis shows the same ordering in achievable SNR~\cite{SWAN}. For ISAC, a hybrid segment selection and multiplexing (HSSM) mode is especially useful: only a subset of segments is activated, but the selected segments retain independent feed weights. HSSM can avoid wasting power on unfavorable segments while preserving multi-user and sensing beamforming capability~\cite{SegPinchISAC}.

Segmentation does not imply that more segments are always better. Short segments reduce guided-wave loss and improve spatial locality, but dividing a fixed number of antennas among too many segments can reduce per-segment aperture gain. Additional feeds, switches, and calibration states also increase cost. The segment number, segment length, and active protocol must therefore be selected jointly with the communication and sensing requirements.

\section{Solution Pathways for PASS-Enabled ISAC}
\label{section:4}

\subsection{Physics-Consistent Modeling and Metric Design}

A workable PASS-enabled ISAC solution starts from the right physical model. The end-to-end response is not just a free-space beamforming channel. It is the cascade of guided-wave propagation from the feed to the selected pinching antennas and free-space propagation from those antennas to users or targets. This means feed location and radiation coupling should be modeled together with user and target geometry~\cite{PA}. In-waveguide attenuation and feed placement~\cite{DualFedPA}, as well as segment boundaries~\cite{SWAN}, should also be counted. Otherwise, the optimizer may choose a PA position that looks attractive in free space but is actually inefficient once guided-wave loss is counted.

The second modeling step is metric design. PASS is especially sensitive to this choice, as different ISAC goals prefer different geometries. If the sensing function is coverage oriented, illumination power or sensing SNR may be an adequate constraint in power-position design~\cite{follow} or segmented SWAN-ISAC control~\cite{SegPinchISAC}. If the goal is localization, the formulation should move to Fisher information, CRLB~\cite{PAISACCRLB}, or BCRB for multicast-sensing settings~\cite{PAMulticast}. If target echo strength changes with view angle, outage or detection reliability becomes more meaningful than deterministic illumination~\cite{TargetDiversity}. In other words, the sensing metric should be chosen before the optimization method, since it determines what kind of geometry PASS should create.

\subsection{Joint Spatial-Signal Optimization}

Joint spatial-signal optimization couples continuous PA positions, power, and beamforming weights with discrete guide activation, segment selection, protocol modes, and candidate locations. Moving a PA changes path loss and phase; activating a guide or segment changes the independent feed resources. Joint beamforming~\cite{PASSjoint}, WDMA~\cite{WDMA}, and segmented ISAC~\cite{SegPinchISAC} therefore require coordinated decisions.

Alternating optimization, matching, penalties, and successive convex approximation offer interpretable methods for joint beamforming~\cite{PASSjoint} and WDMA~\cite{WDMA}. Learning-based controllers can support rapid adaptation, provided they enforce power-position constraints~\cite{follow} and segmented protocol constraints~\cite{SegPinchISAC}.

\begin{figure*}[t]
	\centering
	\figureframe{\includegraphics[width=0.7\textwidth]{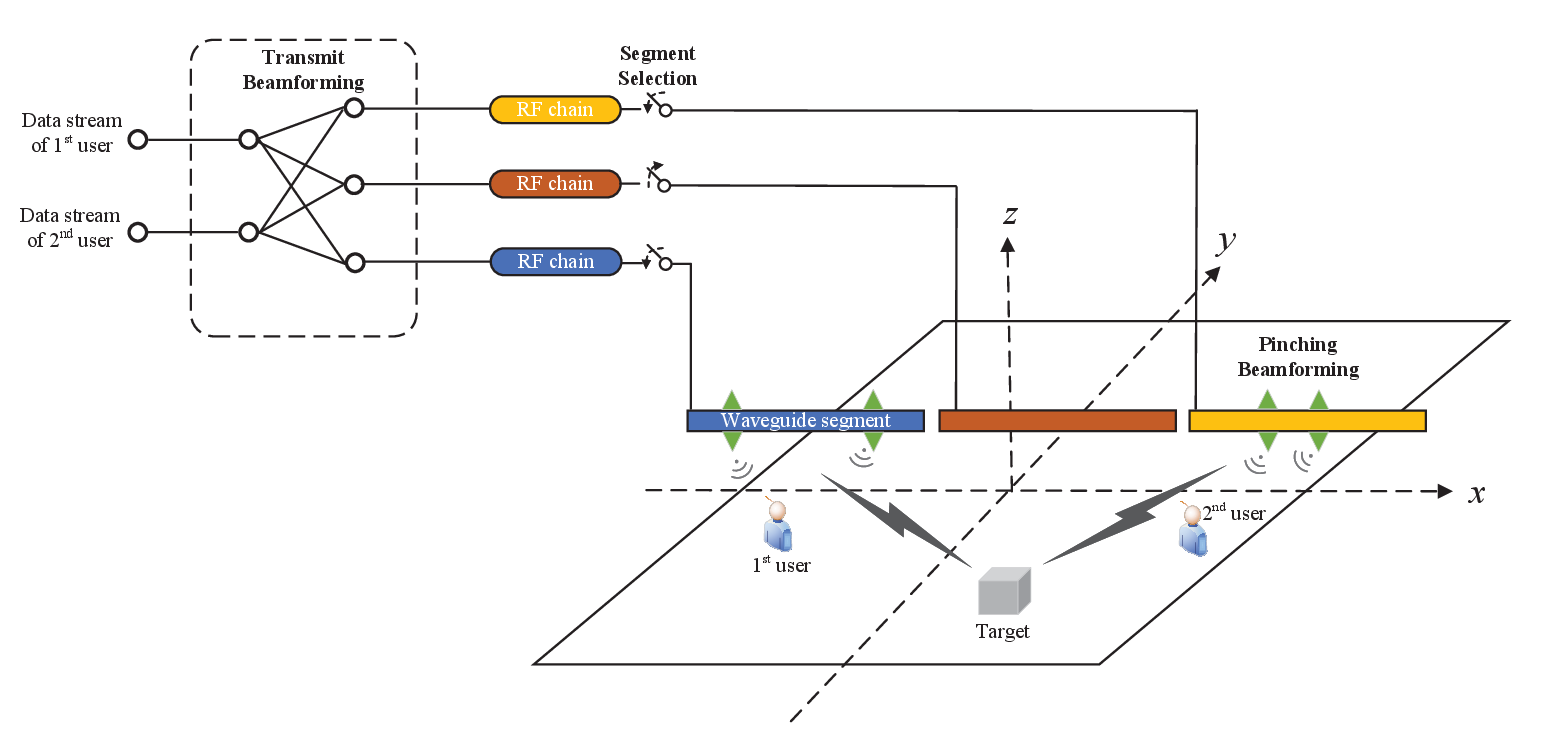}}
	\caption{Illustration of a SWAN-ISAC scenario.}
	\label{fig3}
\end{figure*}

\subsection{State Acquisition for a Reconfigurable ISAC Aperture}

Optimization alone is not enough, given that PASS changes the aperture while the system is operating. A practical solution therefore needs joint state acquisition for communication and sensing. Estimating an independent channel coefficient for every possible PA position is inefficient and often ill conditioned. A better pathway is to exploit position metadata, waveguide transfer structure, and spatial sparsity. Position-aware learning architectures such as PAMoE and PAformer follow this direction for PASS channel estimation~\cite{PASSCE}. Noncoherent differential spatial modulation provides a complementary route for reducing dependence on instantaneous CSI when PA positions change frequently~\cite{DSMPA}. Localization-oriented estimation can also use the multi-view geometry of PASS to infer user and scatterer locations rather than only channel coefficients~\cite{PASSLoc}.

For PASS-enabled ISAC, this estimation layer should become state inference rather than conventional CSI acquisition. Communication pilots can reveal user paths, sensing echoes can update target geometry, and calibrated waveguide responses can interpolate channels or sensing views at inactive positions. Segment-level calibration is particularly appealing, since only the changed segment may need retraining after reconfiguration. The practical goal is to provide the optimizer with a compact, uncertainty-aware state that remains valid even when geometry changes.

\subsection{Hierarchical Online Control with Topology Stability}

A final solution ingredient is hierarchical control. PASS-enabled ISAC naturally operates on multiple timescales: layout and segmentation are slow variables, segment or waveguide activation is an intermediate variable, and beamforming or scheduling is a fast variable. Updating all of them every slot is unnecessary and can create excessive switching. A more scalable pathway is to use a slow controller for topology-changing actions and a fast controller for signal-level adaptation. Candidate pruning, action hysteresis, and change penalties are especially important in segmented systems, as they prevent the controller from chasing marginal gains at the cost of unstable hardware reconfiguration~\cite{SegPinchISAC}.

\section{Case Study: Segmented-Waveguide PASS for ISAC}
\label{section:5}

\subsection{SWAN-ISAC Architecture and Joint Design}

A base station controls $M$ short waveguide segments deployed above a service area. Each segment contains position-adjustable PAs and has an accessible feed, enabling the selected segments to carry independently weighted signals as shown in Fig. \ref{fig3}. The representative simulation considers three segments with ten PAs per segment, six communication users and one sensing target in a $50\,\mathrm{m}\times60\,\mathrm{m}$ area. The carrier frequency is $28$~GHz, the total power budget is $100$~W, the receiver noise power is $-90$~dBm, and the waveguide height is $5$~m.

The architecture adopts HSSM. Segment selection removes unfavorable or unnecessary feeds, while multiplexing over the selected segments preserves multi-user beamforming capability. Within each active segment, pinching beamforming controls the PA positions and the guided/free-space phase relationship. The resulting design maximizes the communication sum rate subject to a minimum target-illumination requirement, a total power constraint, segment-selection constraints, and minimum spacing between PAs. This is a representative mixed-control problem since segment selection is discrete, whereas PA positions and transmit beamforming are continuous.

\subsection{Segment-Hysteresis Reinforcement Learning}

The proposed segment-hysteresis reinforcement-learning (SHRL) controller is built on an actor-critic policy. Its state contains communication and sensing channel information together with the previous PA positions and segment assignment. The action contains updated PA positions, raw segment-selection logits, and beamforming parameters. The reward is the communication sum rate penalized when the target illumination falls below its required threshold.

The distinctive component is a hysteresis gate applied to the structural action. At each decision step, a proposed segment assignment is either accepted with an update probability or replaced by the previous assignment. This simple memory mechanism separates two behaviors that should not evolve at the same speed: continuous beamforming and PA positions can be refined frequently, whereas the segment topology changes more conservatively. The gate reduces abrupt remapping and the non-stationarity seen by the actor-critic learner. A periodic-update baseline (SPRL) also reduces switching, but it cannot adapt the update timing to the learned policy.

\subsection{Results and Design Lessons}

\begin{figure*}[t]
\centering
\figureframe{%
\begin{tikzpicture}
\node[anchor=south west,inner sep=0] (img) at (0,0)
  {\includegraphics[width=0.8\textwidth]{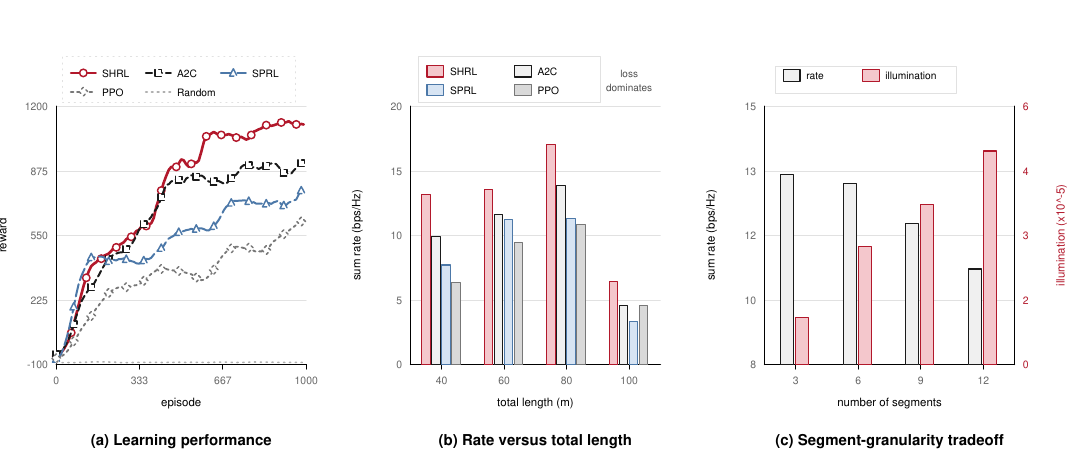}};
\begin{scope}[x={(img.south east)},y={(img.north west)}]
\draw[line width=0.45pt] (-0.005,0.035) rectangle (0.325,0.945);
\draw[line width=0.45pt] (0.325,0.035) rectangle (0.650,0.945);
\draw[line width=0.45pt] (0.650,0.035) rectangle (1.000,0.945);
\end{scope}
\end{tikzpicture}%
}
\caption{Case-study results for SWAN-ISAC: (a) learning performance of SHRL and baselines; (b) communication rate versus total waveguide length for $M=3$; and (c) communication-sensing tradeoff versus segment number for a fixed $40$~m total length.}
\label{fig:case_results}
\end{figure*}

Fig.~\ref{fig:case_results}(a) summarizes the learning curves from different algorithms. Over the final ten stored evaluations, SHRL obtains an average reward of approximately $1115$, compared with $880$ for A2C, $734$ for periodic segment updating, $588$ for PPO, and $-89$ for random control. The result supports a specific conclusion: when one action changes the physical topology, stabilizing that action can be more important than using a more sophisticated generic policy update.

The architecture results in Fig.~\ref{fig:case_results}(b) and (c) reveal two non-monotonic tradeoffs. With three segments, increasing the total waveguide length from $40$ to $80$~m raises the SHRL communication rate from $13.15$ to $17.07$~bps/Hz since the larger aperture improves spatial matching. Extending it to $100$~m reduces the rate to $6.41$~bps/Hz as guided-wave loss and inefficient aperture extension dominate. At a fixed total length of $40$~m, increasing the segment number from three to twelve raises the reported sensing-illumination metric from $1.09\times10^{-5}$ to $4.96\times10^{-5}$, but lowers the communication rate from $13.15$ to $10.59$~bps/Hz. Shorter segments improve target illumination by reducing PA-to-feed loss, whereas fewer PAs per segment weaken communication array gain.

These results motivate joint selection of segment granularity and service objectives, with topology updates controlled separately from continuous signal adaptation.

The case study also has boundaries. It uses line-of-sight spherical-wave channels and illumination power; it does not yet cover target-parameter estimation, clutter rejection, switching latency, or measured waveguide hardware. The values therefore compare algorithms rather than provide universal benchmarks. Next validation should add measured insertion loss, calibration error, CRLB- or detection-based sensing metrics, and segment-switching statistics.

\section{Directions for Future Research}
\label{section:6}

\subsection{Closed-Loop Transmit-Receive PASS-Enabled ISAC}

Future PASS-enabled ISAC should jointly design illumination, echo reception, clutter suppression, target-parameter estimation, and feedback-driven reconfiguration. Multicast-sensing PASS offers richer joint service objectives~\cite{PAMulticast}, while segmentation and local feeds can support transmit-receive coordination~\cite{SWAN}. For multiple targets, dynamic scatterers, and simultaneous uplink services, transmit and receive geometries must co-evolve. Validation should therefore assess estimation or detection performance under changing geometry, beyond transmit-side illumination alone.

\subsection{Measurement-Calibrated Digital Twins and Experimental Platforms}

Measurement-calibrated digital twins should connect algorithms to hardware behavior. Ideal attenuation, coupling, switching delay, and calibration assumptions can conceal errors that degrade both communication and sensing. A twin should track waveguide transfer functions, segment health, blockage, echo history, and environmental uncertainty using online measurements. Experimental platforms should measure insertion loss, coupling drift, reconfiguration latency, and sensing degradation under calibration error. Together, these tools can move costly exploration offline while supporting calibration and control with measured evidence.

\subsection{Trustworthy and Scalable PASS Infrastructure Control}

Scaling PASS-enabled ISAC requires control over many PA locations, waveguides, segments, and mobile users or targets. Controllers should budget switching frequency, preserve minimum sensing service during adaptation, and tolerate incomplete state information. Robust resource allocation under imperfect CSI already motivates modeling position uncertainty~\cite{RobustPA}. Future work should extend this to calibration drift, blockage, and malicious geometry updates. Protecting the control plane, reporting sensing confidence, and limiting location-data exposure should accompany performance optimization, with reliability evaluated under faults and adversarial perturbations.

\section{Conclusion}
\label{section:7}

PASS enables ISAC to adapt radiation geometry through waveguide-based apertures. Single, multiple, and segmented waveguides offer different feed resources and protocol flexibility, requiring coordinated modeling, optimization, state acquisition, and control. The SWAN-ISAC case study shows that guided-wave loss and per-segment aperture gain create competing communication and sensing benefits. Practical deployment will require measured calibration, stable reconfiguration, and receive-side sensing validation to translate this geometric flexibility into reliable service.

\bibliographystyle{IEEEtran}
\bibliography{ref}

\end{document}